\documentclass[a4paper]{jpconf}
\usepackage{amsmath,amssymb,bbm}
\usepackage{amsfonts}
\usepackage{graphicx}
\usepackage{pgf}

\newlength{\figurewidth}
\newcommand\Schr{Schr\"odinger}

\newcommand{\tpl}{t_\mathrm{Pl}}
\newcommand{\lpl}{\ell_\mathrm{Pl}}
\newcommand{\mpl}{m_\mathrm{Pl}}
\newcommand{\epl}{E_\mathrm{Pl}}
\newcommand{\ldb}{\lambda_\mathrm{dB}}
\newcommand{\mean}{\mathbb{E}}
\newcommand{\xv}{\mathbf{x}}
\newcommand{\yv}{\mathbf{y}}

\begin{document}
\title{Planck length challenges non-relativistic quantum mechanics  
     of large masses}

\author{Lajos Di\'osi}

\address{Wigner Research Centre for Physics, H-1525 Budapest 114, P.O.Box 49, Hungary}

\ead{diosi.lajos@wigner.mta.hu}

\begin{abstract}
With the simplest proof ever, we justify the significance of 
quantum-gravity in non-relativistic quantum mechanics together with the related 
theories and experiments. 
Since the de Broglie wave length is inverse proportional to the mass, it would descend
towards and below the Planck scale $10^{-33}$ cm for large masses even at slow
non-relativistic motion. The tricky relationship between gravity and quantum 
mechanics ---well-known in the relativistic case--- shows up in non-relativistic motion of 
massive objects. Hence the gravity-related modification of their 
\Schr~equation is \emph{mandatory}. We also recall the option of an autonomous 
Newtonian quantum-gravity, a theory parametrized by $\hbar$ and $G$. 
On cancellation of $c$ from
the Newtonian limit of Planck scale metric fluctuations is given a new hint.
\end{abstract}

\section{Introduction}
Quantum-gravitational limitations of our very notion of space-time were conjectured
by Bronstein in 1936 \cite{Bro36}, the same issue was famously characterized  by 
Wheeler's  foamy structure of space-time \cite{Whe62}
In the back-ground of quantized matter, the classical notion of space-time continuum 
can not be maintained at short distances. In the vicinity of the Planck length 
\begin{equation}\label{lpl}
\lpl =\sqrt{\frac{\hbar G}{c^3}}=1.6\times10^{-33} cm,  
\end{equation}
the quantization of space-time would become unavoidable according to mainstream 
views. Efforts to construct a satisfactory theory of quantum-gravity have 
reached partial successes only.  
There is no consensual model of  distances about $\lpl$ \cite{Gar95,Hos13}.  

Direct observation of distances like $10^{-33}$ cm  
would require incredible high precision --- beyond imaginations. Their indirect test 
would be possible around the extreme high Planck energy $\epl=1.2\times 10^{19}$ GeV 
per particle, which existed right after the Big Bang only. At low energies, Planck scale 
effects are too small to be testable. That's the mainstream wisdom at least. 

``... it certainly is not a viable option to simply wait until the Planck scale 
becomes accessible experimentally in order to have an indication on
how to proceed theoretically, as this will require a long time'' --- says 
the recent review Ref. \cite{Basetal17} arguing for low-energy gravity-related 
decoherence theories and experiments. There are many discussions of 
quantum-gravity related effects that might occur at non-relativistically low energies,
cf., e.g. \cite{Hos13,Basetal17,Ameetal05,Piketal12,Howetal18}.

But, what is the reason that the extreme-relativistic Planck scale effects, 
suppressed by a factor $10^{-28}$ at low energies (a number used in \cite{Ameetal05}),
may amplify (or accumulate) to the level of laboratory testability?
Here we are going to show the trivial answer lies in non-relativistic 
quantized motion of  massive objects, and the argument is no doubt
simpler than any earlier ones.

\section{Non-relativistic quantized motion reaches Planck length}
Let us consider and object of mass $m$ and compare its de Broglie wave length $\ldb$
with the Planck length $\lpl$: 
\vskip-28pt
{\Large 
\begin{equation}
\underbrace{\lambda_\mathrm{deBroglie}}_{\frac{2\pi\hbar}{mv}\sqrt{1-v^2/c^2}}
\sim~
\underbrace{\ell_\mathrm{Planck}}_{\sqrt{\frac{\hbar G~}{\!c^3}}}\nonumber
\end{equation}
}
\hspace{-5pt}where $v$ is the object's velocity. Now, $\ldb$ can sink to $\lpl$ in two ways.
First: velocity $v$ approaches $c$. Obviously, the required energy
is $\epl$, too high to occur apart from the Big Bang. Second: mass $m$ 
grows macroscopic keeping $v\ll c$. That's the way we go.

\subsection{Periods of the non-relativistic plane wave}
Construct the non-relativistic time-dependent plane wave $\Psi(x,t)$, at
velocity $v\ll c$, along the coordinate $x$:
\begin{equation}
\begin{array}{ll}\Psi(x,t)&=\exp\left(-iEt/\hbar+ipx/\hbar\right)\\
                                             &=\exp\left(-i\frac{mv^2}{2\hbar}t+i\frac{mv}{\hbar}x\right)\\
                                             &=\exp\left(-2\pi i\frac{t}{\tau}+2\pi i\frac{x}{\lambda}\right),
\end{array}
\end{equation}              
where $E,p$ are the energy and momentum, respectively. 
The plane wave is periodic both in time and space, with the corresponding periods $\tau$ 
and $\lambda=\ldb$:
\begin{subequations}
\begin{eqnarray}
\tau&=&(4\pi\hbar/mv^2),\\
\ldb&=&(2\pi\hbar/mv).
\end{eqnarray}
\end{subequations}
These periods should be compared to the Planck time $\tpl$ and length $\lpl$:
\begin{subequations}
\begin{eqnarray}\tpl&=&\sqrt{\hbar G/c^5}\sim10^{-43} s,\\
                               \lpl&=&\sqrt{\hbar G/c^3}\sim10^{-33} cm.
\end{eqnarray}
\end{subequations}
$\Psi(x,t)$ is legitimate non-relativistic wave function as long as its periods are much
longer than the Planck time and length, respectively:
$\tau\gg\tpl$ and $\ldb\gg\lpl$, discussed similarly in \cite{Gao17}.
That's the case for atomic masses $m$, where
$\tau/\tpl\sim10^{18}$ and $\ldb/\lpl \sim 10^{18}$.
But much larger masses will push the structure of $\Psi(x,t)$ towards the Planck scales, 
such role of the mass is what our work aims to emphasize.

\subsection{How large should the mass be?}
Masses $M$, requested for a Planck scale structure of $\Psi(x,t)$ non-relativistically, 
are much greater than the Planck mass $\mpl=2.18\times10^{-5}$ g. 
The time-period $\tau$ would sink to the Planck time $\tpl$ if the mass is
\begin{equation}
M\vert_{\tau=\tpl}=\frac{c^2}{v^2}\mpl\sim\frac{c^2}{v^2}10^{-5} g,
\end{equation}
whereas the spatial period $\ldb$ drops down to $\lpl$ provided
\begin{equation}
M\vert_{\ldb=\lpl}= \frac{c}{v}\mpl\sim\frac{c}{v}10^{-5} g.
\end{equation}
Remember that we keep $v<<c$. Interestingly, the condition $\tau=\tpl$ 
is equivalent with $Mv^2/2=\epl/2$, i.e., the center-of-mass wave function temporal structure
approaches the Planck scale $\lpl$ provided the non-relativistic kinetic energy approaches,
due to the large $M$, the Planck energy $\epl$ (see related observation in \cite{Gao17}). 
Btw, the incoherent preparation of macroscopic objects, unlike of elementary  particles,
of Planckian kinetic energy $\epl$ would not be an issue.

But,  for the spatial structure, the bell rings earlier at much smaller mass $M$ and 
kinetic energy  $Mv^2/2=(v/c)\epl$. Through examples, we illustrate that the system can be
perfectly non-relativistic, though apparently not available by state-of-the-art
coherent preparation of massive objects.  
Our first example is mass $M=10$ g at velocity $v=10$ km/s, yielding  
$\ldb=4.2\times 10^{-33}$ cm which is of the order of $\lpl$!
Of course, the single plane wave is not necessarily illegitimate in itself since its
Planck scale structure may be the artifact of the chosen frame of
reference. So, we consider the superposition of two opposite plane waves of
the mass $M=10$ g:
\begin{equation}\label{cat}
\vert \!+\!10km\!/\!s\rangle+\vert \!-\!10km\!/\!s\rangle,
\end{equation}
which is a \Schr~cat in momentum. Its wave function
has a Planck scale spatial structure in any reference frame hence
it is not legitimate.

Our second example is a solid mass $M=10$ kg, with an elastic 
vibrational mode of frequency $\omega=100$ kHz. If the amplitude is
$a=0.01$ cm, then $\ldb$ oscillates between infinity and  
$\ldb=(2\pi\hbar\!/\!ma\omega)=4.2\times 10^{-33}$ cm.
Again, standard \Schr~equation becomes
illegitimate for this massive (macroscopic) non-relativistic oscillator.

\subsection{How can tiny Planckian effects accumulate for massive objects?}
As we mentioned, in non-relativistic atomic quantum systems the Planck scale effects
are suppressed by an incredible small factor guessed deliberately  by $10^{-28}$
which is the ratio of typical atomic energies (eV's) to $\epl\sim10^{19}$ GeV
\cite{Ameetal05}.  We can illustrate how such small effects can accumulate
in macroscopic massive systems. Our example is a minimum toy model of the 
uncertainty $u$ of the classical spatial continuum at the Planckian distances.  
We suppose the replacement 
\begin{equation}
x\Rightarrow x+u
\end{equation}  
where $u$ is a global uncertainty of the coordinate $x$.  It can be quantum or 
just classically random, we choose the latter option for its simplicity. Let $u$
be a Gaussian random number of squared spread $\mean u^2=\lpl^2$. Consider the
many-body wave function of a macroscopic object. Due to the random 
uncertainty $u$, it should be replaced randomly as 
\begin{equation}\label{PsixPsixu}
\Psi(x_1,x_2,\dots,x_{10^{23}})\Rightarrow\Psi(x_1+u,x_2+u,\dots,x_{10^{23}}+u).
\end{equation}
The shifts $u\sim\lpl$ are irrelevant for the non-relativistic individual constituents, 
but their effects accumulate for the center-of-mass. The mechanism of accumulation
is best seen in momentum representation of Eq. (\ref{PsixPsixu}):
\begin{equation}
\widetilde\Psi(p_1,\!p_2,\!\dots,\!p_{10^{23}})\Rightarrow
\exp\left(\frac{i}{\hbar}u(\underbrace{p_1\!+\!p_2\!+\!\dots\!+\!p_{10^{23}}}_{P})\right)\widetilde\Psi(p_1,\!p_2,\!\dots,\!p_{10^{23}}).
\end{equation}
Observe the appearance of the phase $u/\hbar$ times the total
momentum $P$ of the center-of-mass motion of the macroscopic object. 
$P$ is the ``large'' number to compensate the the smallness of $u$.
We can derive closed expression for the influence of the Planckian 
coordinate uncertainty $u$ on the center-of-mass density matrix if we take the
average over $u$:
\begin{equation}
\rho(P,P^\prime)=\mean\exp\!\left(\frac{i}{\hbar}u(P-P^\prime\right)\!\!\rho(P,P^\prime)
=\exp\!\left(\!-\frac{\lpl^2}{2\hbar^2}(P-P^\prime)^2\right)\!\!\rho(P,P^\prime).
\end{equation}
In our toy model, the Planck scale coordinate uncertainty decoheres the center-of-mass momentum
superpositions of massive objects.  As to our \Schr~cat (\ref{cat}), 
the interference term $\rho(mv,-mv)$ is suppressed by the factor
\begin{equation}
\exp\left(-\frac{\lpl^2}{2\hbar^2}(2mv)^2\right)\approx e^{-1.2}.
\end{equation}
The cat (\ref{cat}) becomes partially decohered. 
The atom-wise ignorable Planck scale uncertainty grew testable indeed 
non-relativistically in the massive degree of freedom. 
Yet a consistent theory (advancing our minimum toy model) of Planck scale 
uncertainties should either eliminate the illegitimate cat states 
or make them legitimate. 

To improve the phenomenology of our minimum toy model, the unnatural
globality and constancy of $u$ should be replaced by some correlated 
structure of local space-time randomnesses. They result in much stronger 
decoherence effects whose significance extends for masses $M$ much
below $\mpl$ towards mesoscopic masses. Cat states (\ref{cat}) become
eliminated at fast decoherence rates.
Numerous different choices of space-time fluctuations have been summarized 
recently in Ref. \cite{KwoHog16}, we discuss the particular DP-model below.

\section{Newtonian quantum gravity: $\mathbf{\hbar+G}$, no $\mathbf{c}$}
So far we argued how quantum-gravity, relativistic by its definition,  should 
influence non-relativistic quantum mechanics. The extreme-relativistic quantum-gravity 
mechanism must have its footprints on non-relativistic quantum mechanics
of massive objects. Now, this opens a remarkable perspective: what if velocity of
light $c$ cancels from the said footprints? Then there would be an autonomous
theory of Newtonian quantum-gravity, coined so by \cite{DioLuk87}. 
And, apparently, this shall be the case.
A first indication was the non-relativistic limit (cf. footprint) of quantum-gravity
in the semi-classical approximation which led to the Newton-\Schr~
equation, containing $G$ and $\hbar$ only, and predicting a regime of 
significance for quantum dynamics of meso- and macroscopic masses \cite{Dio84,Pen96}.
The other instance is the DP-model \cite{Pen96,Dio87} of gravity-related decoherence
and wave function collapse in massive degrees of freedom, parametrized 
again by $G$ and $\hbar$ only. It is based on a
conjectured ultimate uncertainty of space-time but neither the relationship
to the Planck length $\lpl$ (sketched in \cite{Pen16}) nor the cancellation of $c$ are
explained in concrete forms. 

The question itself was explicitly asked in Ref. \cite{Dio18}: 
whether the space-time Newtonian unpredictabilities/fluctuations 
$\emph{\'a la}$ DP \cite{Dio87} are the non-relativistic limit of the Planckian's?
A possible, affirmative, answer may come from Unruh's intuitive quantization of space-time
\cite{Unr84}, as mentioned in \cite{Dio87,TilDio17,Dio18}. Below we show a direct affirmative
derivation, still starting from a phenomenology of Planckian fluctuations.
Let us represent the uncertainty around the Minkowski metric by introducing
a perturbative conformal factor $1+h$ (at $\vert h\vert\ll1$) where the random fluctuations $h$ are
proportional to $\lpl$. We choose the formal expression of relativistic covariance:
\begin{equation}\label{hh}
\mean h(x) h(y)=\mathrm{const.}\times\frac{\lpl^2}{(2\pi)^4}\int e^{-ik(x-y)}\frac{\theta(-k^2)}{-k^2}dk,
\end{equation}
ignoring the issues of regularizing $\theta(-k^2)/k^2$.
Now we write $h$ into the form $h=2\Phi/c^2$ anticipating that $\Phi$ plays the
role of a (random) Newton potential for non-relativistic matter.  
In the limit $c\rightarrow\infty$, the good thing about the formal relativistic 
correlator (\ref{hh}) is that it leads to a convergent correlator for $\Phi$:
\begin{equation}
\mean \Phi(t,\xv)\Phi(s,\yv)=\mathrm{const.}\times\frac{\hbar G}{\vert \xv-\yv\vert}\delta(t-s).
\end{equation}
The conform uncertainties, proportional to $\lpl$, have become equivalent non-relativistically 
with the (Newtonian) space-time uncertainties proposed for the DP-model \cite{Dio87}.
The way  $c$ has gone is being demonstrated directly, tentatively though,
for the first time. 

\section{Summary}
We wished relaxing the belief, based mainly on quantum-field-theory considerations,
that the significance of quantum-gravity is restricted for the regime of extreme high energies. 
We simply argue that the Planck length $\lpl$ invokes a \emph{mandatory}  
---gravity-related, yet to be specified--- modification of non-relativistic quantum mechanics of massive 
(macroscopic) degrees of freedom because their de Broglie wave lengths can 
descend into the vicinity of $\lpl$ non-relativistically.   The modification
depends on the yet also unknown behavior of space-time at the Planckian scale.

No matter whether space-time is classical or quantized, some ultimate
statistical or quantum uncertainties (fluctuations) are attributed to the 
background space-time, to result in some particular decoherence of massive 
degrees of freedom. We explained the accumulation of the tiny 
extreme-relativistic effects in massive degrees of freedom. Such non-relativistic 
footprints of the Planck scale fluctuations do, in general, depend on $G$, 
$\hbar$, and $c$. However, the footprints according to the DP-model are 
fully non-relativistic, the light velocity $c$ cancels from them. 
A mechanism of this cancellation has been outlined.

\ack
The author thanks the National Research Development and Innovation Office of Hungary 
Projects Nos. 2017-1.2.1-NKP-2017-00001 and K12435, and the EU COST Action CA15220 
for support.


\section*{References}
\begin{thebibliography}{99}
\bibitem{Bro36} Bronstein M 1936 {\it Phys. Z. Sowjetunion} {\bf 9} 140 
\bibitem{Whe62} Wheeler J A 1962 {\it Geometrodynamics} (New York: Academic Press)
\bibitem{Gar95} Garay L J 1995 {\it Int. J. Mod. Phys. A} {\bf 10} 145
\bibitem{Hos13}  Hossenfelder S 2013 {\it Living Rev. Rel.} {\bf 16} 2
\bibitem{Basetal17} Bassi A, Grossardt A and Ulbricht H 2017 
{\it Class. Quantum Grav.} {\bf 34} 193002
\bibitem{Ameetal05} Amelino-Camelia G, L\"ammerzahl C, Macias A and M\"uller H 
2005 {\it AIP Conf. Proc.} {\bf 758} 30
\bibitem{Piketal12} Pikovski I, Vanner M R, Aspelmeyer M, Kim M S and Brukner \v C 2012
{\it Nat. Phys.} {\it 8}
\bibitem{Howetal18} Howl R, Hackerm\"uler L, Bruschi D E and 
Fuentes I 2018 {\it Advances in Physics: X} {\bf 3}, 1383184
\bibitem{Gao17} Gao S 2017  {\it Meaning of the Wave Function} 
(Cambridge, UK: Cambridge University Press)
\bibitem{KwoHog16} Kwon O and Hogan C J 2016 {\it Class. Quantum Grav.} {\bf 33} 105004
\bibitem{DioLuk87} Di\'osi L and  Luk\'acs B 1987 {\it Annln. Phys.} {\bf 499} 488
\bibitem{Dio84} Di\'osi L 1984 {\it Phys. Lett. A} {\bf 105} 199
\bibitem{Pen96} Penrose R 1996 {\it Gen. Relativ. Gravit.} {\bf 28} 581
\bibitem{Dio87} Di\'osi L 1987 {\it Phys. Lett. A} {\bf 120} 377
\bibitem{Pen16} Penrose R 2016 {\it Fashion, Faith, and Fantasy in the New Physics of the Universe} 
(Princeton, NJ: Princeton University Press) p 365
\bibitem{Unr84} Unruh W G 1984 {\it Quantum Theory of Gravity}, ed S M
Christensen (Bristol, UK: Adam Hilger Ltd, Bristol) p 234
\bibitem{TilDio17} Tilloy A and Di\'osi L 2017 {\it Phys. Rev. D} {\bf 96} 104045
\bibitem{Dio18} Di\'osi L 2018 {\it Entropy} {\bf 20} 496
\end{thebibliography}
\end{document}